\documentclass[a4paper]{article}
\usepackage{spconf,amsmath,graphicx,subfigure}
\usepackage{booktabs}
\usepackage{multirow}
\usepackage{threeparttable}

\title{Key-Sparse Transformer for Multimodal Speech Emotion Recognition}

\name{Weidong Chen$^{\star}$ \qquad Xiaofen Xing$^{\star}$ \qquad Xiangmin Xu$^{\star}$\sthanks{Corresponding author. Email: xmxu@scut.edu.cn}  \qquad Jichen Yang$^{\star}$ \qquad Jianxin Pang$^{\dagger}$}

\address{$^{\star}$ School of Electronic and Information Engineering, South China University of Technology, China
  \\ $^{\dagger}$ UBTECH Robotics Corp, China}
  
\begin{document}

\maketitle
\begin{abstract}
Speech emotion recognition is a challenging research topic that plays a critical role in human-computer interaction. Multimodal inputs further improve the performance as more emotional information is used. 
However, existing studies learn all the information in the sample while only a small portion of it is about emotion. 
The redundant information will become noises and limit the system performance.
In this paper, a key-sparse Transformer is proposed for efficient emotion recognition by focusing more on emotion related information. 
The proposed method is evaluated on the IEMOCAP and LSSED. Experimental results show that the proposed method achieves better performance than the state-of-the-art approaches.
\end{abstract}


\section{Introduction}
\label{sec:intro}

Speech emotion recognition (SER) is fast becoming a key instrument in human-computer interaction (HCI) \cite{10.1145/3129340}. 
SER also sheds new light on autism and the elderly care and so on, which are collectively referred to healthcare \cite{6026823}. 
For example, the people who suffer from severe speech and language disorder have difficulty expressing their emotions. 
An emotion recognition system can help to treat the patients and improve their emotional communication skills. 

Speech is multimodal as it contains text information by its nature.
Latest researches \cite{Pan2020, huang2021makes} have also proved that multimodal methods outperform the uni-modal methods. 
Consequently, multimodal SER has been a hot research topic in recent years. For example, 
\emph{Yoon et al.} \cite{8639583} use dual recurrent neural networks to  combine the information from audio and text.
In the same way, \emph{Krishna et al.} \cite{N.2020} use raw audio waveform as audio features and GloVe word embeddings as text features for multimodal learning. 
Moreover, \emph{Peri et al.} \cite{add_citation} combine audio and video information and utilize multitask setting for emotion recognition.
In this paper, we use both audio and text information for SER.

Pre-trained Self Supervised Learning (SSL) has 
made great success 
in many fields such as natural language processing \cite{2018arXiv181004805D, 2019arXiv190711692L}
and speech recognition \cite{DBLP:conf/interspeech/SchneiderBCA19}. 
Meanwhile, recent works \cite{9206016, Siriwardhana2020} that use SSL model have obtained promising results in SER. 
Nowadays, wav2vec \cite{DBLP:conf/interspeech/SchneiderBCA19} and RoBERTa \cite{2019arXiv190711692L} are the most commonly used pre-trained SSL models in the literature.
Thus, in this paper, we use them to extract audio and text embeddings, respectively.

\begin{figure}[t]
  \centering
  \includegraphics[width=\linewidth]{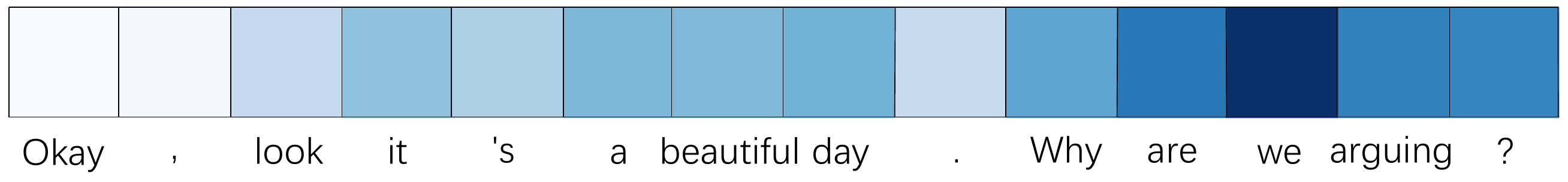}
  \caption{The attention weights of the utterance ``OK, look it's a beautiful day. Why are we arguing?" in vanilla Transformer. Darker colors represent larger weights.} 
  \label{fig:attention_weights}
\end{figure}

Inspired by the attention mechanism, Transformer \cite{10.5555/3295222.3295349}, which is outstanding in modeling long sequence, is proposed and has achieved great success in natural language processing \cite{9206016}. 
Meanwhile, several Transformer based architectures have been introduced for SER. 
\emph{Tarantino et al.}~\cite{Tarantino2019} use global windowing system in Transformer to capture deep relationships within the utterance. Moreover,
\emph{Huang et al.} \cite{9053762} use Transformer to fuse different modalities for sentiment analysis. 
In this paper, we use Transformer as our basic structure to implement emotion recognition.


However, few works have paid attention to that not all the information in audio or text is related to emotion. For example, considering a text \emph{``Okay, look it's a beautiful day. Why are we arguing?"} in IEMOCAP \cite{IEMOCAP}, the attention weights in vanilla Transformer are shown in Figure~\ref{fig:attention_weights}. 
We can see that the attention weights in Transformer are assigned to all the words.
However, words \emph{``beautiful"} and \emph{``arguing"} contain the majority of emotional information in this sentence.
And the words that are not related to emotion such as \emph{``it"}, \emph{``a"} and \emph{``look"}, are unnecessary for SER task and become noises, leading to the limitation of system performance. To address this issue, we propose a novel method, named key-sparse Transformer (KS-Transformer), to judge the importance of each word or speech frame in the sample and help the model focus more on the emotion related information. 
Based on KS-Transformer, we further design a cascaded cross-attention block to fuse different modalities with high efficiency.

The contributions of this paper can be summarized as follows:
\begin{itemize}
\item We propose KS-Transformer to judge the importance of each frame or word
that helps the model focus more on the emotional information.
Based on KS-Transformer, we further design a cascaded cross-attention block to achieve interaction between different modalities.
\item We evaluate the proposed method on IEMOCAP and LSSED, and demonstrate that it achieves better results than the existing state-of-the-art approaches.
\end{itemize}

\section{Proposed Method}

\begin{figure}[ht]
  \centering
  \includegraphics[width=0.65\linewidth]{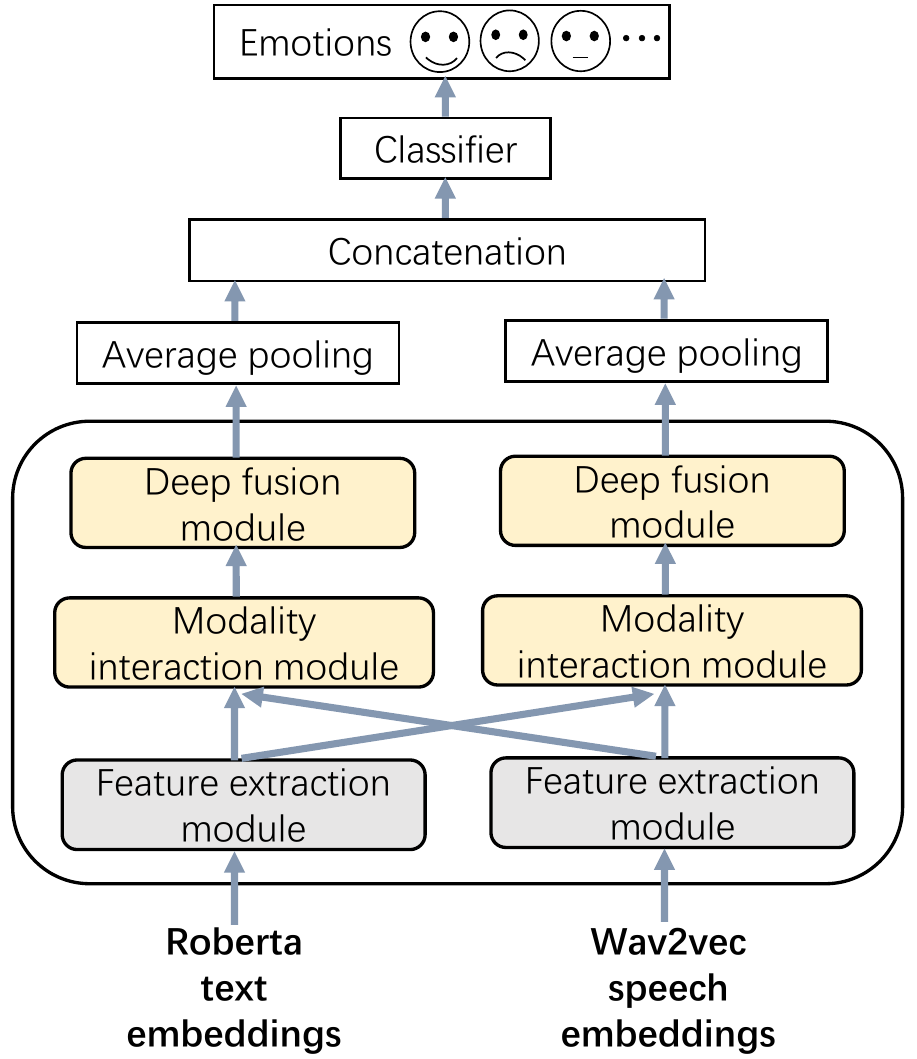}
  \caption{Overview structure of the proposed model.}
  \label{fig:Overview_structure}
\end{figure}

The proposed model, as shown in Figure~\ref{fig:Overview_structure}, mainly consists of three modules. In which, feature extraction module is used to learn the input features, modality interaction module is used for learning interactive information and deep fusion module aims to further combine the information from audio and text. 
Specifically, the first module (gray parts) is based on vanilla Transformer and the last two modules (yellow parts) are based on KS-Transformer.
More details will be introduced in the following subsections.

\subsection{Key-Sparse Transformer}

\subsubsection{Vanilla Transformer}
Vanilla Transformer consists of encoder and decoder originally. In this paper, we use Transformer to represent the encoder part, 
since it is the one needed for the implementation of our proposed architecture.
The inputs of Transformer are divided into \emph{\textbf{Q}}, \emph{\textbf{K}} and \emph{\textbf{V}}, which consist of Query, Key and Value vectors, respectively. 
The attention mechanism in vanilla Transformer is depicted as follows:
\begin{equation}
  \emph{\textbf{W}} = softmax(\frac{\emph{\textbf{QK}}^{T}}{\sqrt{d_{Q}}})
  \label{eq1}
\end{equation}
\begin{equation}
  \emph{\textbf{attn}} = \emph{\textbf{W}}\times \emph{\textbf{V}}
  \label{eq2}
\end{equation}
where d$_{Q}$ is the dimension of the Query vector, \emph{\textbf{W}} is the weight matrix and \emph{\textbf{attn}} is the attention output. For multi-head attention mechanism, we combine the attention outputs
from all the heads. More details can be found in \cite{10.5555/3295222.3295349}.

\subsubsection{Key-Sparse attention mechanism}
\begin{figure}[ht]
  \centering
  \includegraphics[scale=0.4]{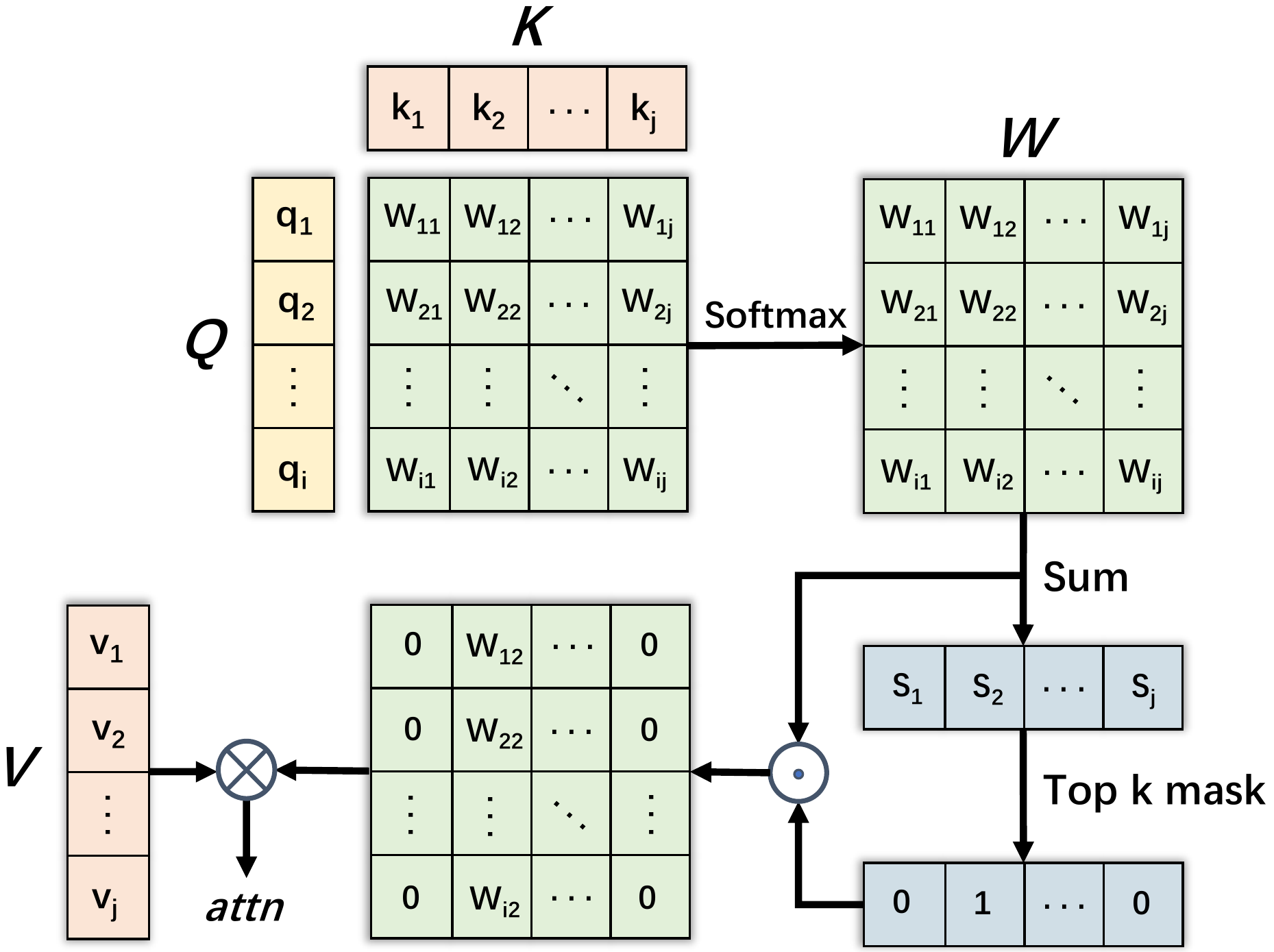}
  \caption{The key-sparse attention in KS-Transformer. In which, softmax and summation are performed on each row and column, respectively. $\odot$ and $\otimes$ represent position-wise multiplication and  matrix multiplication, respectively.}
  \label{fig:sparse_attention}
\end{figure}

The key-sparse Transformer aims to find the emotional information automatically.
Assume the number of Query vectors in \emph{\textbf{Q}} is \emph{i} while that of Key vectors in \emph{\textbf{K}} is \emph{j}, the key-sparse attention mechanism is illustrated in Figure~\ref{fig:sparse_attention}. 
It should be noted that \emph{\textbf{K}} and \emph{\textbf{V}} are always the same in Transformer.

Key-sparse attention mechanism, which is used in KS-Transformer, is capable of judging the importance of each speech frame or word automatically.
As shown in Figure~\ref{fig:sparse_attention}, the weight matrix \emph{\textbf{W}} is obtained by multiplying \emph{\textbf{Q}} and \emph{\textbf{K}}, and each row in \emph{\textbf{W}} are the weights of Value vectors in \emph{\textbf{V}}.
As a Value vector represents a frame in audio or a word in text, we add up all the weights of the same Value vector and the summation is used as a discriminator for the importance of the speech frame or word in the sample.
We select \emph{k} Value vectors with \emph{top-k} largest summation and keep their attention weights in weight matrix unchanged while the others are reset to zero. This operation makes the weight matrix from dense to sparse and reduces the redundancy, that's why we call the Transformer used here as KS-Transformer. The \emph{top-k} mask is calculated by Equation \ref{eq4}.
\begin{equation}
  \emph{\textbf{M}}_\textbf{z}=
  \begin{cases}
    0      & \text{if}\ s_z < threshold    \\ 
    1      & \text{if}\ s_z \geq threshold
  \end{cases}
  \label{eq4}
\end{equation}
where \emph{threshold} is the \emph{k$^{th}$} largest summation and $z\in[1,j]$.

\subsection{Modality interaction module}

\begin{figure}[ht]
  \centering
  \includegraphics[scale=0.5]{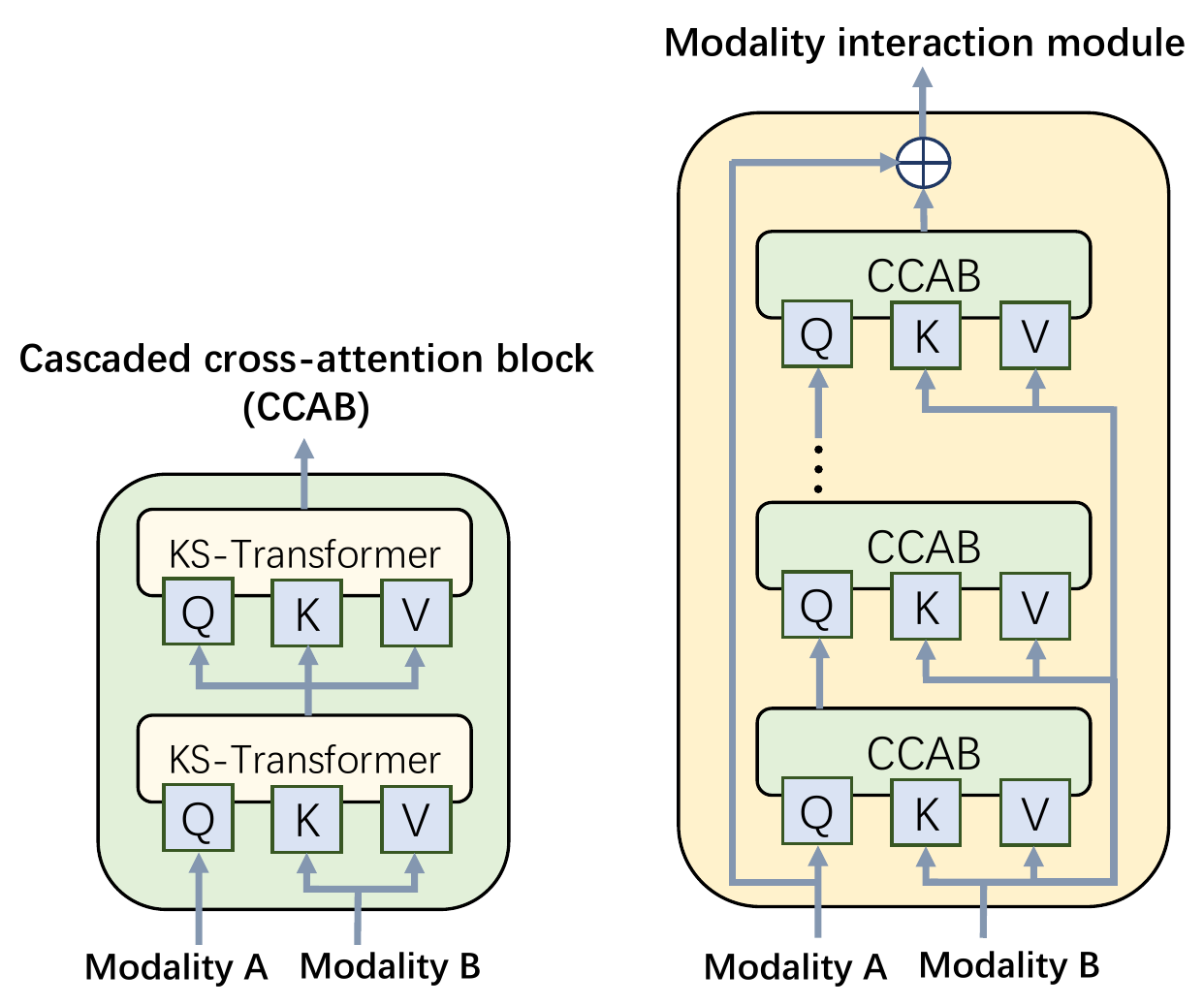}
  \caption{The details of CCAB (left) and modality interaction module (right).}
  \label{fig:second_module}
\end{figure}

Because modality interaction module is based on cascaded cross-attention block (CCAB), we introduce CCAB's structure first.
As shown in the left part of Figure~\ref{fig:second_module},
CCAB is a cascade of two KS-Transformers, in which, the first KS-Transformer creates \emph{\textbf{Q}} from modality \emph{A} and \emph{\textbf{K}}, \emph{\textbf{V}} from modality \emph{B}. With this special input method, the key-sparse attention mechanism will find out the most relevant part in \emph{B} for \emph{A} and produce an output which has combined \emph{A} with \emph{B} information. 
Since the emotional information between different modalities is often complementary \cite{Pan2020, LiuPengfei2020, Chen2020}, neither \emph{A} nor \emph{B} can represent the accurate emotion. Therefore, 
the second KS-Transformer in CCAB takes the fused features as input and considers the information from both modality \emph{A} and modality \emph{B} when applying key-sparse attention. Benefited from CCAB, \emph{A} and \emph{B} are fused more comprehensively and accurately.

As shown in the right part of Figure~\ref{fig:second_module}, modality interaction module consists of a stack of CCABs, wherein the later CCAB takes the output of the former CCAB as \emph{\textbf{Q}} input while \emph{\textbf{K}} and \emph{\textbf{V}} inputs are always from modality \emph{B}.
That the information from \emph{B} goes through one CCAB is regarded as one interaction because the information from \emph{B} had flowed into \emph{A} by the key-sparse attention. 
More than one CCAB are applied for multiple times interactions.  A skip connection is utilized for the features' stability.


\subsection{Deep fusion module}

Most researches take the fused features to predict emotions after the interaction \cite{Siriwardhana2020, Mittal_Bhattacharya_Chandra_Bera_Manocha_2020}. 
However, we argue that the fused features maybe not the best and can be deep fused to further improve the system performance.
In detail, deep fusion module consists of several KS-Transformers, 
in which, they take the fused features as input and utilize key-sparse attention to enhance the interaction between audio and text and implement deep fusion.


\section{Experiments}

\subsection{Database introduction}

IEMOCAP contains five sessions, every of which has one male and one female speaker, respectively. 
To stay consistent with the previous works~\cite{N.2020, LiuPengfei2020, Chen2020}, 
we use 5,531 utterances from four emotions: angry, neutral, happy (\& excited) and sad.
We conduct experiments in leave-one-session-out cross-validation strategy. 

LSSED \cite{9414542} is a new released large-scale English speech emotion dataset, which has data collected from 820 subjects and contains 147,025 samples. Consistent with \cite{9414542}, we use four emotion categories, including angry, neutral, happy and sad. For each emotion class, its associated samples are randomly split into train/development/test in ratio of 7/1/2, respectively.
Every experiment is run for 10 times to avoid randomness, and the averaged result is used as the final accuracy.

\subsection{Experimental setup}

The pre-trained wav2vec and RoBERTa are available online\footnote{https://github.com/pytorch/fairseq}.
The max lengths of the audio and text feature sequence are set to 460 and 20, respectively.
SGD optimizer with a learning rate of 5$\times$10$^{-4}$ on IEMOCAP and 1$\times$10$^{-4}$ on LSSED is applied to optimize the model.
The learning rate drops to 50$\%$ of the original every 30 epochs.
Dropout with \emph{p} = 0.5 is utilized to alleviate over-fitting. The batch size is 32. 

Feature extraction module is used to learn the input features, which are extracted from pre-trained SSL models, aims to obtain suitable features for SER task. For modeling rich contexts, this module is based on vanilla Transformer.
\emph{\textbf{Q}}, \emph{\textbf{K}} and \emph{\textbf{V}} inputs here are the same, which is known as self-attention \cite{10.5555/3295222.3295349}. 
The number of vanilla Transformers in feature extraction module is 5 and the number of KS-Transformers in deep fusion module is 2. 
Eight attention heads are used in multi-head attention. 
The number of CCABs used in modality interaction module will be discussed later. 

\subsection{Experimental results and analysis}

\subsubsection{Key-sparse attention analysis}

\begin{figure}[t]
\centering
\subfigure[Vanilla Transformer]{
\begin{minipage}[t]{\linewidth}
\centering
\includegraphics[width=0.9\linewidth]{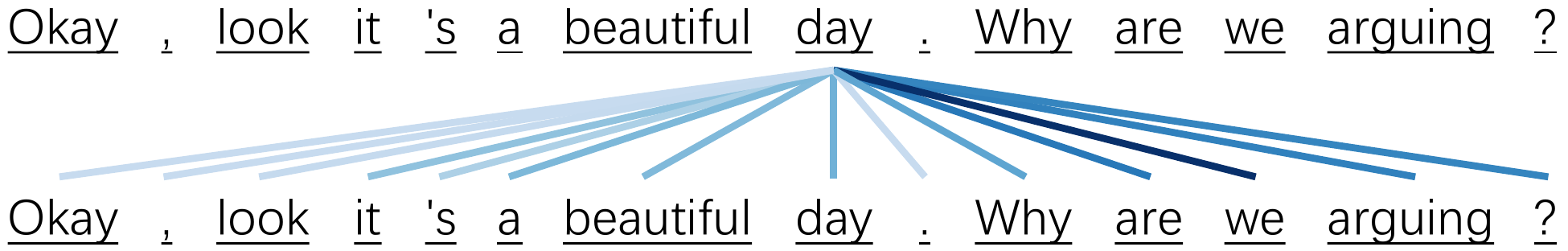}
\end{minipage}
}%
\\
\subfigure[KS-Transformer]{
\begin{minipage}[t]{\linewidth}
\centering
\includegraphics[width=0.9\linewidth]{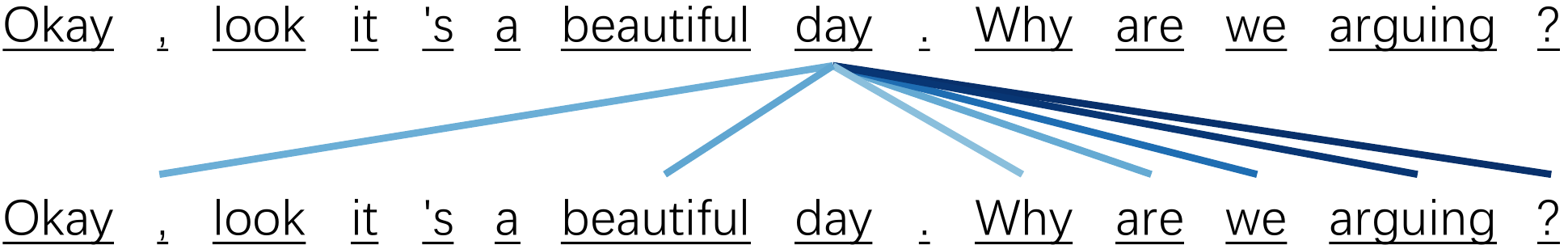}
\end{minipage}
}%
\centering
\caption{Visualization of the attention weights.}
\label{fig:Key_sparse_analysis}
\end{figure}

To demonstrate the effectiveness of the key-sparse attention, we consider a sample in IEMOCAP and compare the attention weights in vanilla Transformer and KS-Transformer by visualization. 
As shown in Figure~\ref{fig:Key_sparse_analysis}, the vanilla Transformer takes note of all the words, including the noisy words which are not related to emotion, and trends to over-fitting. 
However, the KS-Transformer makes the connections from dense to sparse, which is able to ignore most of the noises and focus more on the emotional information.
Meanwhile, the sparsity in KS-Transformer can reduces the complexity in the model and alleviates over-fitting.

\begin{figure}[t]
\centering
\subfigure[IEMOCAP]{
\begin{minipage}[t]{0.5\linewidth}
\centering
\includegraphics[width=1.0\linewidth]{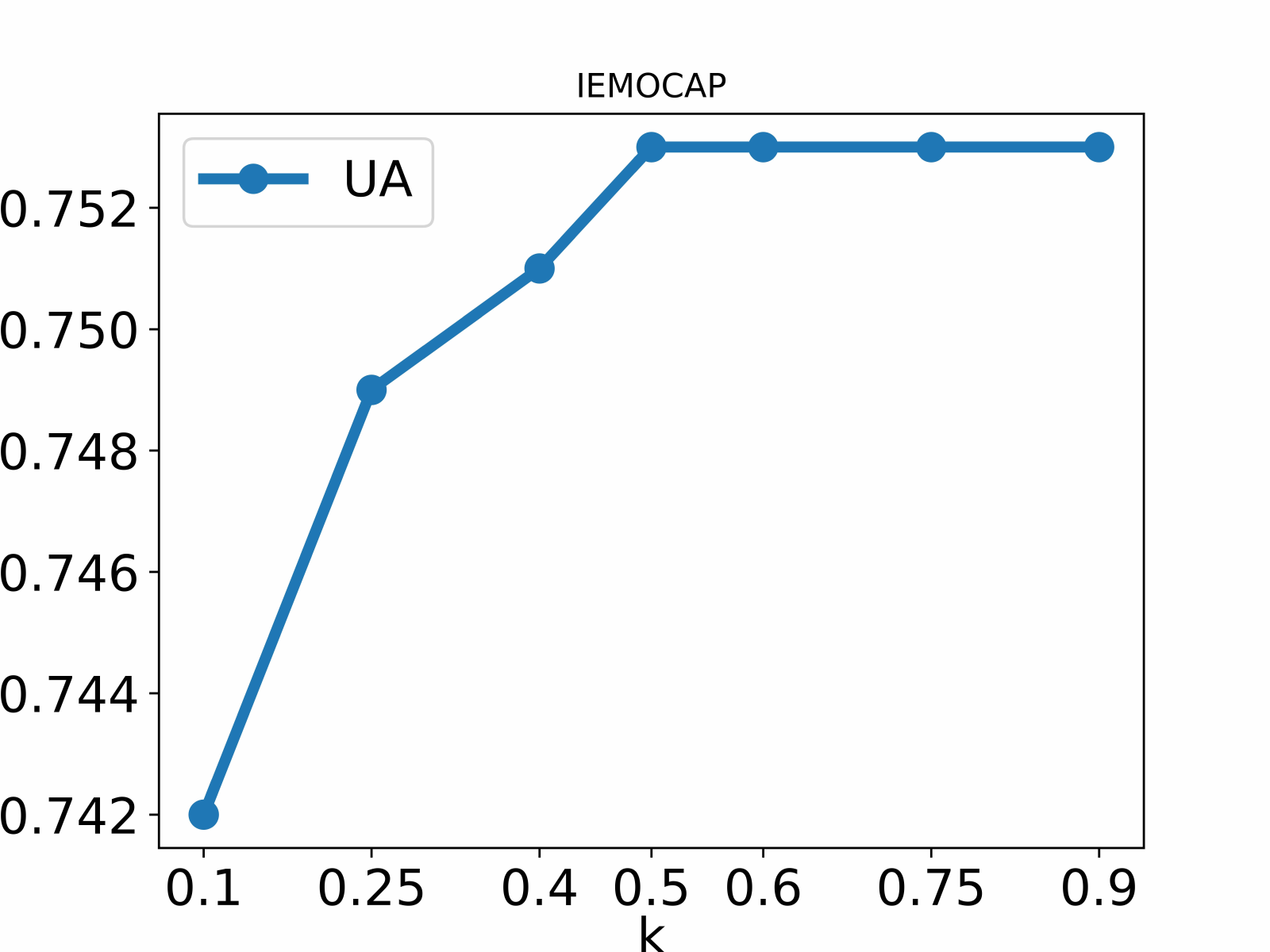}
\end{minipage}%
}%
\subfigure[LSSED]{
\begin{minipage}[t]{0.5\linewidth}
\centering
\includegraphics[width=1.0\linewidth]{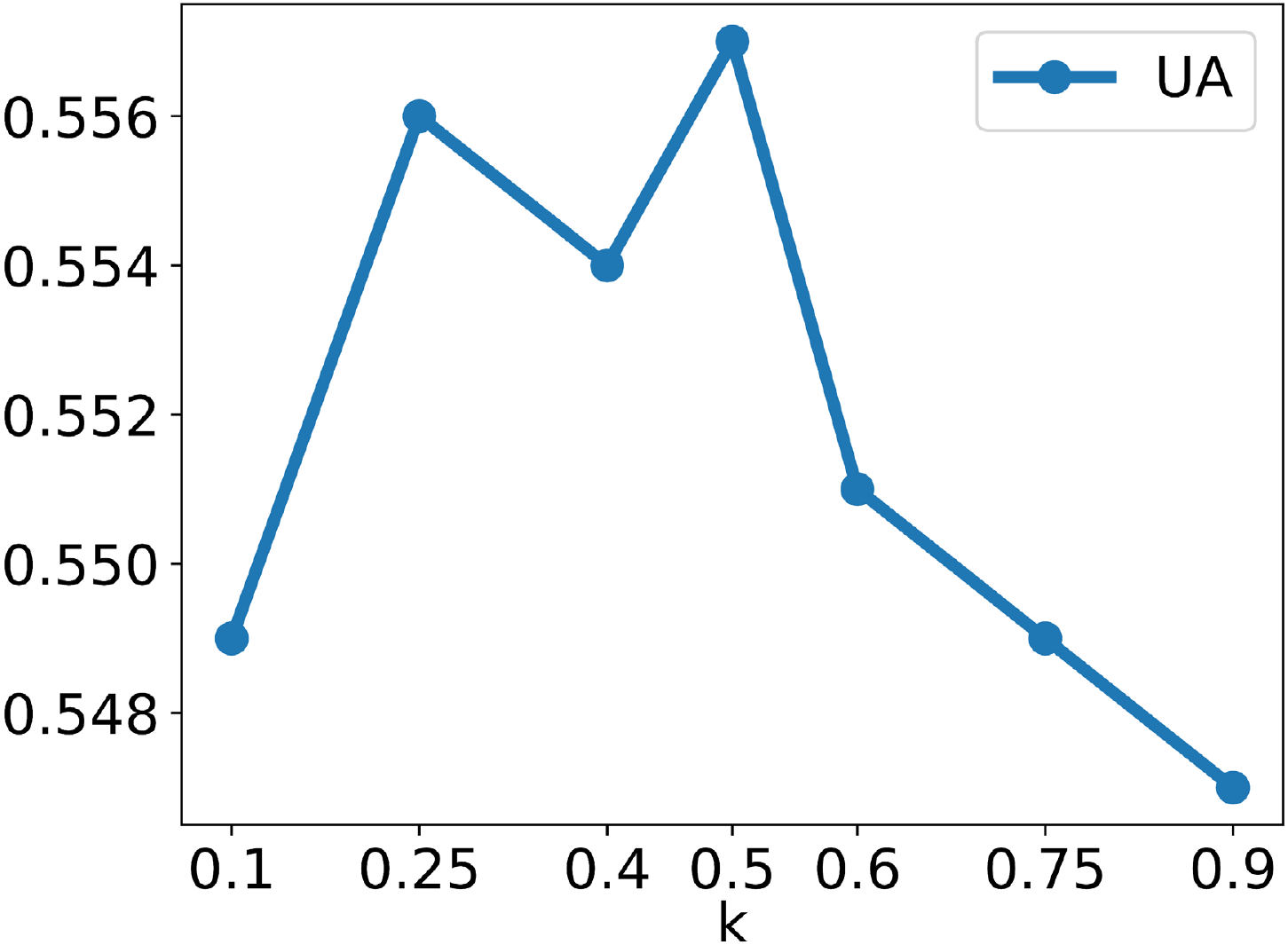}
\end{minipage}%
}%
\centering
\caption{Effects of hyperparameter k.}
\label{fig:k_curve}
\end{figure}

To explore the optimal sparsity in KS-Transformer, we vary \emph{k} from 0.1 to 0.9. The larger \emph{k} we set, the less attention weights are reset to zero and less sparsity we have. 
Because LSSED suffers from sample imbalance, we use unweighted accuracy (UA) as criterion. 
The results are shown in Figure~\ref{fig:k_curve}.

Since IEMOCAP is a relatively small corpus, the model is prone to over-fitting when \emph{k} is larger than 0.5, causing the UA scores to remain constant. However, on the large-scale dataset LSSED, a significant drop is appeared when \emph{k} is larger than 0.5 because of the redundant information. 
In contrast, when \emph{k} is smaller than 0.5, the model uses too little information and might converge to an unsatisfactory local minimum.
Considering the UA performance curves on IEMOCAP and LSSED corpora, \emph{k} is set to 0.5, which means 50$\%$ attention weights are reset to zero in each KS-Transformer as default.

\subsubsection{Multimodal interaction analysis}


Modality interaction is vital for multimodal system. 
To investigate the effectiveness of the stack of CCABs, we change the number of CCABs used from zero to four, where zero means that the modality interaction module is removed, and the results are shown in Table~\ref{tab:CCAB_number}. Weighted accuracy (WA) and UA are used as criteria. 
It should be noted that the number of CCABs used represents the times of interactions performed.

From Table~\ref{tab:CCAB_number}, we show that the interaction between different modalities is shallow and insufficient when only one CCAB is applied. The performance improves as the number of CCABs increases. 
The best performances are obtained when the number is three,
which confirms the effectiveness of CCAB and 
the necessity of multiple times interactions.

\begin{table}[ht]
  \caption{Performances of different number of CCABs in modality interaction module on IEMOCAP and LSSED.}
  \label{tab:CCAB_number}
  \centering
  \begin{tabular}{ccc|cc}
    \toprule
    \multirow{2}{*}{\textbf{Amount}}   & \multicolumn{2}{c|}{IEMOCAP} & \multicolumn{2}{c}{LSSED} \\   & \textbf{WA}  & \textbf{UA}   & \textbf{WA}    & \textbf{UA}\\
    \midrule
    0               & 0.726                & 0.734            & 0.647            & 0.540           \\
    1               & 0.724                & 0.735            & 0.648            & 0.544           \\
    2               & 0.731                & 0.740            & \textbf{0.651}   & 0.554           \\
    \textbf{3}      & \textbf{0.743}       & \textbf{0.753}   & 0.650            & \textbf{0.557}  \\
    4               & 0.742                & 0.751            & 0.648            & 0.555  \\
    \bottomrule
  \end{tabular}
\end{table}

\subsubsection{Comparison with some known systems}

Table~\ref{tab:comparative_experiments} gives the performance comparison among the proposed method with some known systems on IEMOCAP and LSSED, in which, all the systems apply audio and text as inputs except that PyResNet \cite{9414542} only takes audio information. 

From Table~\ref{tab:comparative_experiments}, it can be observed that our method gives the best WA and UA on IEMOCAP.
Moreover, our method achieves the highest UA on LSSED, where UA is a more important criterion because of the sample imbalance issue.


\begin{table}[h]
 \caption{Comparison results on IEMOCAP and LSSED.}
  \label{tab:comparative_experiments}
  \centering
  \begin{threeparttable}
  \begin{tabular}{ccccc}
    \toprule
    \textbf{Dataset}   &\textbf{Methods} & \textbf{Year}  & \textbf{WA}  & \textbf{UA}   \\
    \midrule
    \multirow{4}{*}{IEMOCAP} & CMA \cite{N.2020}  & 2020  & -    & 0.728 \\
    & STSER \cite{Chen2020}  & 2020 & 0.711          & 0.721  \\
    & GBAN \cite{LiuPengfei2020}  & 2020 & 0.724                & 0.701  \\
    & \textbf{Ours}       & 2021 & \textbf{0.743}       & \textbf{0.753} \\
    \midrule
    \multirow{4}{*}{LSSED} & CMA       & 2020           & 0.616\tnote{\#}           & 0.489\tnote{\#} \\
                              & STSER     & 2020           & \textbf{0.651}\tnote{\#}  & 0.512\tnote{\#} \\
                              & PyResNet  & 2021           & 0.624                     & 0.429           \\
                              & \textbf{Ours}           & 2021           & 0.650                     & \textbf{0.557} \\
    \bottomrule
  \end{tabular}
   \begin{tablenotes}
        \footnotesize
        \item[\#] LSSED is a new released dataset. Author provides these results by reproducing the corresponding methods and training and testing them on LSSED dataset.
    \end{tablenotes}
 \end{threeparttable}
\end{table}

\section{Conclusion}
In this paper, KS-Transformer, using a novel key-sparse attention mechanism, has been proposed for speech emotion recognition. Only the emotion related speech frames in audio or words in text can be considered and assigned with attention weights. And based on KS-Transformer, we further present CCAB to fuse different modalities and achieve deep interaction. Experimental results on IEMOCAP and LSSED demonstrate the effectiveness of KS-Transformer and CCAB. In the future, we plan to combine more modalities to further improve the system performance.

\section{Acknowledgement}
The work is supported in part by the National Natural Science Foundation of China under Grant U1801262, in part by the Key-Area Research and Development Program of Guangdong Province, China, under Grant 2019B010154003, and in part by the Science and Technology Project of Guangzhou under Grant 202103010002.

\bibliographystyle{IEEEbib}
\bibliography{refs}

\begin{thebibliography}{10}

\bibitem{10.1145/3129340}
B.~W. Schuller,
\newblock ``Speech emotion recognition: Two decades in a nutshell, benchmarks,
  and ongoing trends,''
\newblock {\em Commun. ACM}, vol. 61, no. 5, pp. 90–99, Apr. 2018.

\bibitem{6026823}
S.~{Tokuno}, G.~{Tsumatori}, S.~{Shono}, E.~{Takei}, T.~{Yamamoto},
  G.~{Suzuki}, S.~{Mituyoshi}, and M.~{Shimura},
\newblock ``Usage of emotion recognition in military health care,''
\newblock in {\em Defense Science Research Conference and Expo (DSR)}, 2011,
  pp. 1--5.

\bibitem{Pan2020}
Z.~Pan, Z.~Luo, J.~Yang, and H.~Li,
\newblock ``{Multi-Modal Attention for Speech Emotion Recognition},''
\newblock in {\em Proc. Interspeech 2020}, pp. 364--368.

\bibitem{huang2021makes}
Y.~Huang, C.~Du, Z.~Xue, X.~Chen, H.~Zhao, and L.~Huang,
\newblock ``What makes multimodal learning better than single (provably),''
\newblock {\em arXiv preprint arXiv:2106.04538}, 2021.

\bibitem{8639583}
S.~Yoon, S.~Byun, and K.~Jung,
\newblock ``Multimodal speech emotion recognition using audio and text,''
\newblock in {\em 2018 IEEE Spoken Language Technology Workshop (SLT)}, 2018,
  pp. 112--118.

\bibitem{N.2020}
D.~N. Krishna and A.~Patil,
\newblock ``{Multimodal Emotion Recognition Using Cross-Modal Attention and 1D
  Convolutional Neural Networks},''
\newblock in {\em Proc. Interspeech 2020}, pp. 4243--4247.

\bibitem{add_citation}
R.~Peri, S.~Parthasarathy, C.~Bradshaw, and S.~Sundaram,
\newblock ``Disentanglement for audio-visual emotion recognition using
  multitask setup,''
\newblock in {\em IEEE International Conference on Acoustics, Speech and Signal
  Processing (ICASSP)}, 2021, pp. 6344--6348.

\bibitem{2018arXiv181004805D}
J.~{Devlin}, M.~{Chang}, K.~{Lee}, and K.~{Toutanova},
\newblock ``{BERT: Pre-training of Deep Bidirectional Transformers for Language
  Understanding},''
\newblock {\em arXiv preprint arXiv:1810.04805}, 2018.

\bibitem{2019arXiv190711692L}
Y.~{Liu}, M.~{Ott}, N.~{Goyal}, J.~{Du}, M.~{Joshi}, D.~{Chen}, O.~{Levy},
  M.~{Lewis}, L.~{Zettlemoyer}, and V.~{Stoyanov},
\newblock ``{RoBERTa: A Robustly Optimized BERT Pretraining Approach},''
\newblock {\em arXiv preprint arXiv:1907.11692}, 2019.

\bibitem{DBLP:conf/interspeech/SchneiderBCA19}
S.~Schneider, A.~Baevski, R.~Collobert, and M.~Auli,
\newblock ``wav2vec: Unsupervised pre-training for speech recognition.,''
\newblock in {\em Proc. Interspeech 2019}, pp. 3465--3469.

\bibitem{9206016}
S.~{Siriwardhana}, T.~{Kaluarachchi}, M.~{Billinghurst}, and S.~{Nanayakkara},
\newblock ``Multimodal emotion recognition with transformer-based self
  supervised feature fusion,''
\newblock {\em IEEE Access}, vol. 8, pp. 176274--176285, 2020.

\bibitem{Siriwardhana2020}
S.~Siriwardhana, A.~Reis, R.~Weerasekera, and S.~Nanayakkara,
\newblock ``{Jointly Fine-Tuning “BERT-Like” Self Supervised Models to
  Improve Multimodal Speech Emotion Recognition},''
\newblock in {\em Proc. Interspeech 2020}, pp. 3755--3759.

\bibitem{10.5555/3295222.3295349}
A.~Vaswani, N.~Shazeer, N.~Parmar, J.~Uszkoreit, L.~Jones, A.~N. Gomez,
  L.~Kaiser, and I.~Polosukhin,
\newblock ``Attention is all you need,''
\newblock in {\em Proceedings of the 31st International Conference on Neural
  Information Processing Systems}, 2017, pp. 5998--6008.

\bibitem{Tarantino2019}
L.~Tarantino, P.~N. Garner, and A.~Lazaridis,
\newblock ``{Self-Attention for Speech Emotion Recognition},''
\newblock in {\em Proc. Interspeech 2019}, pp. 2578--2582.

\bibitem{9053762}
J.~{Huang}, J.~{Tao}, B.~{Liu}, Z.~{Lian}, and M.~{Niu},
\newblock ``Multimodal transformer fusion for continuous emotion recognition,''
\newblock in {\em IEEE International Conference on Acoustics, Speech and Signal
  Processing (ICASSP)}, 2020, pp. 3507--3511.

\bibitem{IEMOCAP}
C.~Busso, M.~Bulut, C.-C Lee, A.~Kazemzadeh, E.~Mower, S.~Kim, J.~N. Chang,
  S.~Lee, and S.~S. Narayanan,
\newblock ``Iemocap: Interactive emotional dyadic motion capture database,''
\newblock {\em Language Resources and Evaluation}, vol. 42, no. 4, pp.
  335--359, 2008.

\bibitem{LiuPengfei2020}
P.~Liu, K.~Li, and H.~Meng,
\newblock ``{Group Gated Fusion on Attention-Based Bidirectional Alignment for
  Multimodal Emotion Recognition},''
\newblock in {\em Proc. Interspeech 2020}, pp. 379--383.

\bibitem{Chen2020}
M.~Chen and X.~Zhao,
\newblock ``{A Multi-Scale Fusion Framework for Bimodal Speech Emotion
  Recognition},''
\newblock in {\em Proc. Interspeech 2020}, pp. 374--378.

\bibitem{Mittal_Bhattacharya_Chandra_Bera_Manocha_2020}
T.~Mittal, U.~Bhattacharya, R.~Chandra, A.~Bera, and D.~Manocha,
\newblock ``M3er: Multiplicative multimodal emotion recognition using facial,
  textual, and speech cues,''
\newblock {\em Proceedings of the {AAAI} Conference on Artificial
  Intelligence}, vol. 34, no. 02, pp. 1359--1367, Apr. 2020.

\bibitem{9414542}
W.~Fan, X.~Xu, X.~Xing, W.~Chen, and D.~Huang,
\newblock ``Lssed: A large-scale dataset and benchmark for speech emotion
  recognition,''
\newblock in {\em IEEE International Conference on Acoustics, Speech and Signal
  Processing (ICASSP)}, 2021, pp. 641--645.

\end{thebibliography}

\end{document}